\documentclass[twocolumn,nofootinbib]{revtex4}
\usepackage{amsfonts,amsmath,mathrsfs,epsfig,amsbsy,bm,verbatim}
\usepackage{graphicx}

\pdfoutput=1

\newcommand{\beq}{\begin{equation}}
\newcommand{\eeq}{\end{equation}}
\newcommand{\bea}{\begin{eqnarray}}
\newcommand{\eea}{\end{eqnarray}}
\newcommand{\nn}{\nonumber \\}

\begin{document}

\title{Entanglement entropy of compressible holographic matter:\\ loop corrections from bulk fermions}
\author{Brian Swingle, Liza Huijse, and Subir Sachdev}
\affiliation{Department of Physics, Harvard University, Cambridge MA 02138}

\begin{abstract}
Entanglement entropy is a useful probe of compressible quantum matter because it can detect the existence of Fermi surfaces, both of microscopic fermionic degrees of freedom and of ``hidden" gauge charged fermions. Much recent attention has focused on holographic efforts to model strongly interacting compressible matter of interest for condensed matter physics.  We complete the entanglement analysis initiated in Huijse {\em et al.}, Phys. Rev. B 85, 035121 (2012) (arXiv:1112.0573) and Ogawa {\em et al.}, JHEP 1, 125 (2012) (arXiv:1111.1023) using the recent proposal of Faulkner {\em et al.} (arXiv:1307.2892) to analyze the entanglement entropy of the visible fermions which arises from bulk loop corrections.  We find perfect agreement between holographic and field theoretic calculations.
\end{abstract}

\maketitle

\section{Introduction}

Holographic duality \cite{maldacena,polyakov,witten} provides a fundamentally new way to understand the physics of quantum many-body systems (quantum field theories) by mapping them to dual gravitational systems.  Furthermore, when the quantum field theory of interest is strongly coupled and has many degrees of freedom, the corresponding dual gravity picture becomes weakly fluctuating and has few degrees of freedom \cite{magoo,condmat_holo_review}.  Although the simplest holographic systems, namely large $N_c$ non-Abelian gauge theories, are unconventional from a laboratory point of view, universality gives us hope that suitable strongly interacting problems arising in experiment might nevertheless be usefully approximated using holographic machinery.

Within condensed matter physics, experiments on a variety of materials have brought attention to the problem of strongly interacting compressible phases of matter \cite{johnstrangemetals}.  The simplest such phase is the Fermi liquid, which has the property that even with strong bare interactions, there remain renormalized quasiparticle degrees of freedom in terms of which the physics is transparent.  We can also easily obtain compressible phases by breaking the $U(1)$ symmetry (a superfluid) or by breaking translation invariance (a crystal).  The search for and understanding of more exotic examples of compressible phases is a major open problem.

There is a general intuition that any compressible phase which does not break $U(1)$ and translation symmetries must involve fermions in its description (see the discussion in Ref. \cite{2011PhRvD..84b6001H}).  Certainly the Fermi liquid is an example with microsocpic fermion degrees of freedom, but other kinds of compressible phases can occur in systems with only microscopic bosonic degrees of freedom in which there are nevertheless emergent fermionic fields.  These emergent fermions will inevitably couple to some kind of emergent fluctuating gauge field, a situation which is already more analogous to the sorts of gauge theories commonly met in holographic systems.  In this vein, a great deal of effort has been expended modeling compressible matter using holography \cite{2011PhRvD..84b6001H,2012JHEP...01..125O,PhysRevB.85.035121,Lee:2008xf,Faulkner:2009wj,Cubrovic:2009ye,Charmousis:2010zz,Hartnoll:2011dm, Iqbal:2011in,2011PhRvD..84f6009S,Iizuka:2011hg}.  However, one immediately encounters a difficulty: since gauge charged fermions are not gauge invariant operators, one cannot define the Fermi surface in a conventional way, e.g., via singularities in the fermion spectral function.  How then can we detect and characterize such hidden Fermi surfaces?  Or can holography perhaps even give us examples of compressible phases without Fermi surfaces, hidden or otherwise?

Assuming that a Fermi surface of some type must be present, one approach to characterizing it is to use the unique entanglement properties of Fermi surfaces.  Given a spatial region $A$, the entanglement entropy of $A$ (defined below) typically satisfies a boundary law $S(A) \sim |\partial A|$, so if $A$ has linear size $L$, then $S(A)\sim L^{d-1}$ ($d$ is the spatial dimension) \cite{arealaw1,arealaw2}.  Remarkably, conventional Fermi gases and Fermi liquids violate this law by possessing an entropy $S(A) \sim (k_F L)^{d-1} \ln{(k_F L)}$ where $k_F$ is the Fermi momentum \cite{ee_f1,ee_f2,bgs_f1,2006PhRvB..74g3103L,2007JMP....48j2110F,2009arXiv0906.4946H}.  Hence, not only does the entanglement entropy detect the existence of the Fermi surface, it also gives a quantitative measure of the Fermi surface via $k_F$.  The proposal of Refs. \cite{2012JHEP...01..125O,PhysRevB.85.035121} is to use the existence of a logarithmic violation of the boundary law as a signature of a hidden Fermi surface.  This proposal, combined with other evidence (e.g., Friedel oscillations), provides a coupling independent method to detect Fermi surfaces even if they are ``hidden".

In the models we consider below, we are explicitly interested in gauge invariant fermions and associated Fermi liquid-like states coexisting with strongly interacting compressible degrees of freedom \cite{PhysRevLett.90.216403,PhysRevLett.105.151602,2011PhRvD..84b6001H}.  More generally, we conjecture that all compressible phases which do not break a symmetry have an $L^{d-1} \ln{(L)}$ term in their entanglement entropy (perhaps in addition to other terms).  There is already circumstantial evidence for this conjecture in that all understood examples of compressible phases which do not break a symmetry involve Fermi surfaces (see e.g. \cite{nature11732,PhysRevB.84.245127}).  Moreover, the thermal to entanglement entropy crossover analysis of Ref. \cite{2013PhRvB..87d5123S} suggests that systems with thermal entropy $S_{\rm thermal} \sim L^d T^{(d-\theta)/z}$ have such a logarithmic violation of the boundary law when $\theta =d-1$.  Here $z$ is the dynamical exponent, so that the dimensionless variable is $LT^{1/z}$, and $\theta$ is the hyperscaling violation exponent.  $\theta = d-1$ is precisely what arises in system with Fermi surfaces.  If our conjecture is true more generally, it will probably require both monopoles \cite{PhysRevD.86.126003,Faulkner:2012gt} and quantum corrections to the entropy to be included in the analysis.  Here we begin along this path with a simpler analysis of bulk fermions which describe a boundary Fermi liquid state.

In Ref. \cite{PhysRevB.85.035121} an extensive analysis of the entanglement properties of a certain class of holographic compressible phases known as hyperscaling violation geometries was considered.  A crucial part of that analysis was the verification of Luttinger's relation which roughly states that the size of a Fermi surface is related to the density of fermions in a system.  It was shown that bulk charge conservation was equivalent to Luttinger's relation, so that in a system composed of ``visible" gauge invariant fermions and ``hidden" gauge charged fermions, the $k_F$ which appears in the entanglement entropy was controlled just by the ``hidden" charge density.  This is sensible because within the large $N_c$ or classical approximation, the holographic entanglement entropy calculation only detects hidden Fermi surfaces.  It was also shown that the coefficient of the ``hidden" Fermi surface contribution was independent of the presence of ``visible" bulk fermions.  The contribution of ``visible" fermions is suppressed in the $1/N_c$ expansion and represents a loop correction to the classical area formula \cite{holo_ee}.

In this paper we use the recent proposal of Ref. \cite{2013arXiv1307.2892F} to complete the analysis begun in Ref. \cite{PhysRevB.85.035121} by analyzing quantum corrections to the entanglement entropy in systems with bulk fermions at finite density. We show with a simple argument that the ``visible" bulk fermions are, as far as entanglement entropy is concerned, dual to Fermi liquid-like degrees of freedom in the dual field theory, just as was anticipated by previous studies.  Furthermore, because the entanglement entropy of such a Fermi liquid state is known purely on the field theory side, we are able to provide a strong check of the proposal in Ref. \cite{2013arXiv1307.2892F} by showing that the holographic and field theory results agree.

In particular, while there are many subtleties concerning the physical interplay of various cutoffs, e.g., the bulk UV cutoff and the bulk IR/boundary UV cutoff, the logarithmic violation of the boundary law we investigate bypasses these issues.  We need only compare the loop contributions of bulk fermions with and without a finite charge density.  Hence all complications associated with the bulk UV cutoff are irrelevant to the logarithmic violation because it is a bulk infrared effect.  Similarly, as we describe in more detail below, the finite density of bulk fermions also sits near a given value of the radial coordinate and does not explore the full bulk minimal surface.  Hence bulk IR/boundary UV cutoff effects are also irrelevant because the bulk charge is insensitive to the whole bulk minimal surface. Thus our results provide a very clean test of the proposed loop correction in Ref. \cite{2013arXiv1307.2892F}.

The remainder of the paper is structured as follows.  First, we review the proposal of Ref. \cite{2013arXiv1307.2892F} and comment on related evidence.  Second, as a warmup we discuss the case of a holographic Fermi liquid realized in a hard wall geometry.  Third, we discuss the entanglement entropy of bulk fermions in hyperscaling violation geometries.  Finally, we discuss some related problems and future directions.

\section{Quantum corrections to holographic entanglement entropy}

Given a quantum field theory with a tensor product Hilbert space, the density matrix of a subsystem $A$ is given by
\beq
\rho_A  = \text{tr}_B (\rho_{AB})
\eeq
where $\rho_{AB}$ is the state of the whole system, typically a pure ground state or a mixed thermal state.  The entanglement entropy $S(A)$ of $A$ is the von Neumann entropy of $\rho_A$:
\beq
S(A) = -\text{tr}(\rho_A \ln{(\rho_A)}).
\eeq
When $\rho_{AB}$ is a pure state, then $S(A)$ indeed measures entanglement between $A$ and its complement $B$.

To compute the entanglement entropy of a region $A$ in a field theory with a holographic dual in the semiclassical limit, we must construct the bulk minimal surface $W(A)$ with $\partial W = \partial A$ at the boundary of AdS.  The entanglement entropy of $A$ is then
\beq
S(A) = \frac{|W(A)|}{4 G_N}
\eeq
in close analogy with the Bekenstein formula for black hole entropy \cite{PhysRevLett.96.181602}.  We emphasize that this gives only the classical approximation to the entanglement entropy.  This prescription passes many checks and gives sensible answers for the entanglement entropy \cite{2011JHEP...05..036C,PhysRevD.76.106013,Hartman:2013mia,Faulkner:2013yia}.

It is convenient to compute the entanglement entropy using the replica formulation:
\beq
S(A) = -\text{tr}(\rho \ln{(\rho)}) = - \lim_{n\rightarrow 1} \partial_n \text{tr}(\rho_A^n).
\eeq
The partition function which computes $\text{tr}(\rho_A^n)$ has a geometric interpretation in terms of a branched space for each integer $n$.  The analytic continuation to non-integer $n$ is the main difficulty with this approach.

To make some progress, one can consider a Lorentz invariant quantum field theory.  Then when the surface $\partial A$ at $t=0$ (a spacetime codimension two surface) is a Killing horizon, the branched space possesses additional symmetry and one can extend the geometry to non-integer $n$.  For example, consider $d+1$ Minkowski space in which the Killing vector which generates $x$-boosts is
\beq
\xi = x \partial_t + t \partial_x.
\eeq
Using the conventional Minkowski metric $\eta$, the pseudo-norm of this vector is
\beq
\eta(\xi,\xi) = - x^2 + t^2
\eeq
which vanishes on the light cone.  In particular, the point $x=t=0$ is a Killing horizon.  Hence when $A$ is the half-space $x>0$, the replicated geometry possesses an extra symmetry.  In imaginary time, this extra symmetry is simply rotation about the origin.  Now, the importance of this symmetry is that we can use it to define the replicated geometry at non-integer $n$ in a precise way.  To do so, we use the quantum generator $K$ of the boost or imaginary time rotation to write $\text{tr}(\rho_A^n)$ as
\beq
\text{tr}(\rho_A^n) = \text{tr}(e^{- 2\pi n K}).
\eeq
Since $K$ has a direct geometrical meaning, a simple spacetime interpretation of this expression is possible even for non-integer $n$ using (in imaginary time) a spacetime with a conical singularity.

Stationary black hole horizons are also of the type just considered.  For example, in the Schwarzchild black hole, the Killing vector $\partial_t$ has zero pseudo-norm on the event horizon.  This is not surprising since the near horizon region is actually equivalent to the half-space situation just considered.  The procedure above then leads to the result that the entropy is proportional to the area of the black hole horizon.  However, one also learns that quantum corrections due to matter fields can be effectively included by computing the entanglement entropy of the field theory degrees of freedom in the black hole background.

In the context of holographic duality, the minimal area formula \cite{holo_ee} was put forward as a heuristic generalization of the black hole entropy computation to more general surfaces.  The interpretation of the entropy was that it represented entanglement between degrees of freedom in the dual field theory.  Recently, Ref. \cite{2013arXiv1304.4926L} gave an argument for this formula which turns on the idea that for $n\neq 1$, there is an effective spacetime defect $\Sigma$ whose equation of motion requires the bulk area $|\Sigma|$ be minimal.  More recently still, Ref. \cite{2013arXiv1307.2892F} argued that the picture of quantum corrections around black hole geometries also generalized to the minimal surface situation.  See also Ref. \cite{2012arXiv1212.5183B} for a similar earlier argument.

An easy check of this argument is possible when the entangling surface is spherical and the dual field theory is conformal.  As shown in Ref. \cite{2011JHEP...05..036C}, this situation maps to a holographic computation in which the boundary is a hyperbolic space.  Furthermore, in this special case the bulk minimal surface happens to coincide with a stationary black hole horizon, so the black hole machinery immediately implies not only the correctness of the classical area formula but also the validity using the bulk entanglement of matter fields as the leading quantum correction.  In this paper we will give another justification of this proposal in a rather different setting and for arbitrary entangling surface using fermions.

To summarize, the proposal of Ref. \cite{2013arXiv1307.2892F} is that the leading quantum correction comes from the bulk entanglement entropy of all non-metric variables across the minimal surface in the classical background.  There may also be additional terms which can be written as integrals over the bulk minimal surface, for example, a shift in the value of $G_N$ or integrals of the curvature.  As long as these terms associated to the minimal surface cannot change the qualitative behavior of the classical result, then we may just look to the bulk entanglement entropy for new physics.

\section{Warmup: hard wall case}
We wish to study a quantum field theory with a conserved $U(1)$ current $J^\mu$ and a fermion $\Psi$ which carries charge $q$ of this current.  The dual gravitational degrees of freedom are a metric $g$, a gauge field $A$, and a fermion $\psi$ of charge $q$ under $A$.  The dual gravitational action is
\beq
\mathcal{S} = \int d^4 x \sqrt{g} \left[ \frac{R}{2\kappa^2} + \frac{1}{4e^2} F^2 + i \bar{\psi}( \Gamma \cdot D + m) \psi \right]
\eeq
where $D$ is the covariant derivative with couplings to both the gauge field and the spin connection.

Following Ref. \cite{2011PhRvD..84f6009S} we choose the metric to be
\beq
ds^2 = \frac{ - dt^2 + dr^2+ dx^2 + dy^2}{r^2}
\eeq
with the asymptotic boundary (field theory UV) at $r=0$ and a hard wall boundary condition at $r=r_m$.  The gauge field is taken to be $A_t = i h(r)$ and we require that $h(r\rightarrow 0) \rightarrow \mu$ (the chemical potential).  Treating $h(r)$ as a fixed background field to be determined self-consistently, the solution to the equations of motion is determined by the Dirac equation.  Reducing the four component equation to a two component equation as in Ref. \cite{2011PhRvD..84f6009S} and studying energy eigenstates $\chi_{\ell,k}$ with energy $E_{\ell,k}$ and $x$-momentum $k$ we have
\beq
\left( i Y \frac{d}{dr} - X \frac{m}{r} - k Z - q h \right) \chi_{\ell,k}= E_{\ell,k} \chi_{\ell,k} .
\eeq
$X,Y,Z$ are Pauli matrices and $\ell$ labels different discrete energy levels at a given momentum $k$.

As $r\rightarrow 0$, the solutions $\chi$ behave as $\chi \sim r^m$, while at $r=r_m$ we demand the Dirac operator be self-adjoint which requires
\beq
\chi^\dagger_1(r_m) Y \chi_2(r_m) = 0.
\eeq
The $\chi$s are normalized according to
\beq
\int_0^{r_m} dr \chi^\dagger_{\ell,k} \chi_{\ell,k} = 1.
\eeq
This completes the specification of the Dirac problem.

At zero temperature the ground state of the Dirac fermions is obtained by filling up all energy states with $E_{\ell,k} < 0$.  The situation is familiar from the band theory of solids: each label $\ell$ describes a continuous band of states labeled by $k$.  Rotational invariance guarantees that the result depends only on the magnitude of $k$.  All negative energy states within each such band are then filled, and thus as far as the $x-y$ physics is concerned, we simply have a set of partially filled bands.  In the simplest case, some set of bands with $\ell < \ell_0$ will be partially filled with spherical Fermi surfaces and Fermi momenta $k_{F,\ell}$, and all bands with $\ell \geq \ell_0$ will be empty.  Note that all charges and energies are measured relative to the zero chemical potential state.

Given this energy level filling picture, the ground state of the fermions is
\beq
|\mu\rangle = \prod_{\ell <\ell_0} \prod_{|k|<k_{F,\ell}} c^\dagger_{\ell,k} | \text{vac}\rangle.
\eeq
The creation operators $c^\dagger_{\ell,k}$ are defined by the expansion of the field operator as
\beq
\psi(x,t,r) = \sum_\ell \int \frac{d^2 k}{4 \pi^2} \left( \chi_{\ell,k}(r) e^{i k x -i E_{\ell,k} t} c_{\ell,k} + ... \right).
\eeq
To compute the correction to the dual field theory entanglement entropy due to these bulk fermions, we must compute their bulk entanglement entropy across the bulk minimal surface arising from the classical approximation.

\begin{figure}
  \centering
  \includegraphics[width=.48\textwidth]{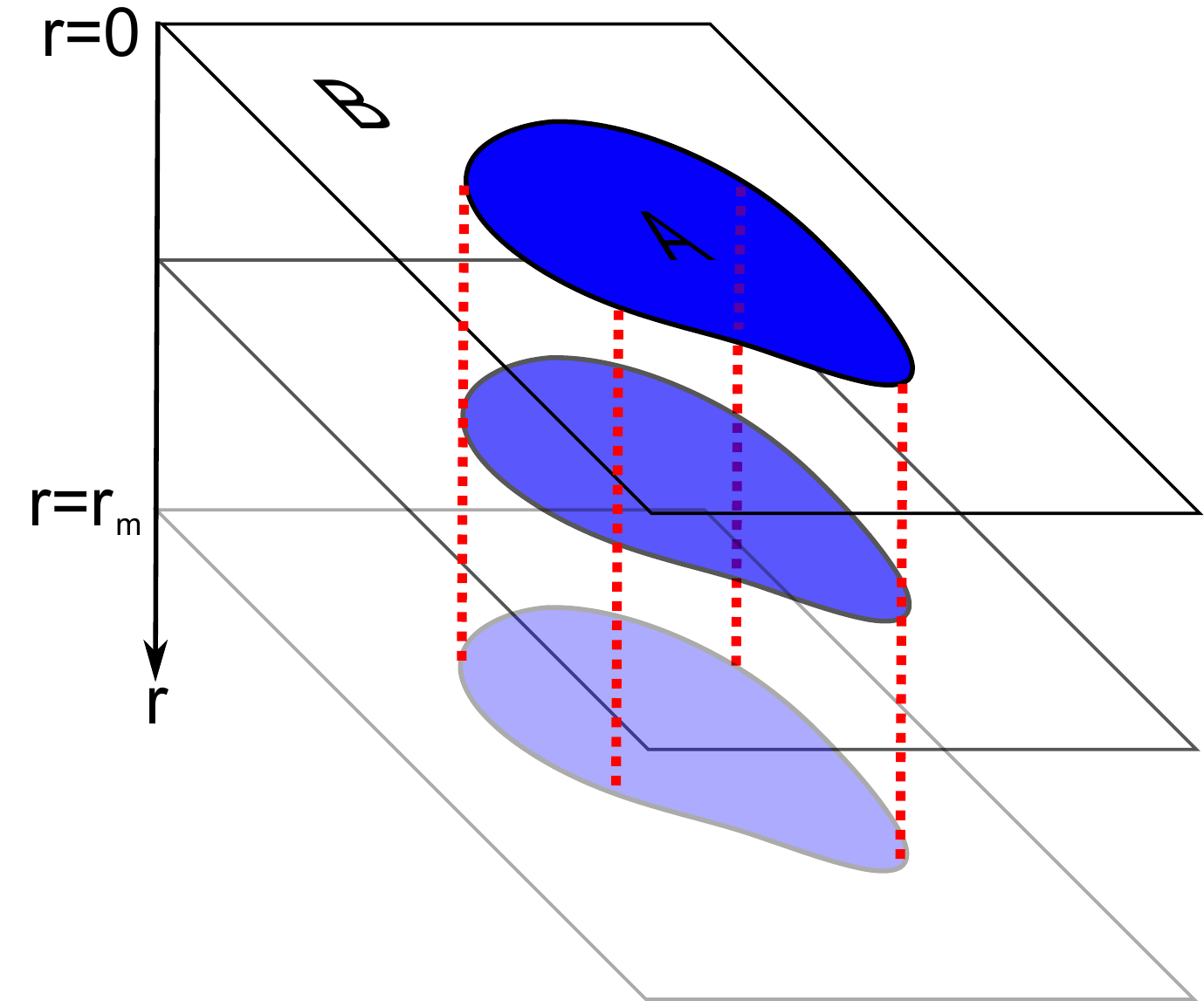}
  \caption{Sketch of the minimal surface when $L \gg r_m$.  The red dashed lines show cuts through the bulk minimal surface. The bulk minimal surface hangs straight down from the boundary to the hard wall so that the shape of the surface with a fixed $r$ plane is independent of $r$.}
  \label{fig:mshw}
\end{figure}

Let us recall the classical approximation to the holographic entanglement entropy.  We choose a region $A$ in the field theory and study the minimal surface $W$ with the property that $\partial W$ at $r=0$ is equal to $\partial A$.  At $r=r_m$ we allow the bulk minimal surface to terminate at the hard wall.  This is sensible to describe a theory without an extensive entropy (as would arise if the minimal surface had to the run along the infrared wall) and mimics the situation obtained in AdS soliton geometries.  Now for a region $A$ of linear size $L$ satisfying $L \ll r_m$, the minimal surface will reduce to a surface in pure AdS and will reproduce the conformal result.  On the other hand, when $L \gg r_m$, the minimal surface will hang approximately straight down to the hard wall and a strict area law will be obeyed:
\beq
S(A) = \frac{|\partial A|}{4 G_N} \int_{r=\epsilon}^{r=r_m} \frac{dr}{r} = \frac{|\partial A|}{4 G_N}\left(\frac{1}{\epsilon} - \frac{1}{r_m} \right).
\eeq

What happens then we turn on a finite chemical potential and consider the bulk fermions?  When $L \ll r_m, \mu^{-1}$ the bulk fermions hardly contribute since the boundary at $r=0$ repels the fermion wavefunctions (because AdS is like a box).  However, when $L \gg r_m, \mu^{-1}$, we find that the fermions make a significant contribution.  Recall that in this case the minimal surface approximately falls straight down to the hard wall.  Crucially, the shape of the minimal surface at a plane of constant $r$ is independent of $r$ and is set by $\partial A$.  Hence when we trade $r$ for the band index $\ell$, each band may be treated as if it is a two dimensional system of fermions in which we are computing the entanglement across a surface of shape $\partial A$.  In more detail, each fermion wavefunction factorizes into a function of $x$ and a function of $r$, so because the bulk surface is approximately independent of $r$ in the relevant region, we may make a basis transformation from $r$ to $\ell$ and hence trace over all $r$ or all $\ell$ to the same effect.

The bulk fermion entanglement entropy, using the Widom formula \cite{ee_f2,bgs_f1}, is
\beq
S_{\text{bulk fermion}} = \sum_{\ell < \ell_0} \frac{k_{F,\ell} |\partial A|}{6 \pi} \ln{(k_{F,\ell} L)},
\eeq
where we have assumed all Fermi surfaces are spherical. Notice that although this correction is formally $1/N_c$ suppressed (no factor of $G_N^{-1}$), for a fixed UV cutoff $\epsilon$ this term eventually dominates the classical contribution.

The retarded bulk fermion two point function has the form
\beq
G^R(k,\omega,r,r') = \sum_{\ell} \frac{\chi^\dagger_{\ell,k}(r) \chi_{\ell,k}(r')}{\omega - E_{\ell,k} + i \delta}.
\eeq
This expression implies that the field theory two point function $G^R_{bdy}$ has the same singularity structure as the bulk fermion two point function.  In particular, the $\chi$s only contribute a finite quasiparticle residue related to their asymptotic value near $z=0$.  Thus the boundary fermion spectral function, obtained from $A \sim \text{Im}(G^R_{bdy})$ has the Fermi liquid form: $A \sim Z \delta(\omega - E_{\ell,k})$.  It has been established on general grounds \cite{PhysRevB.86.035116,PhysRevX.2.011012}, rigorously proven in some models \cite{2012arXiv1209.0769S}, and even checked numerically \cite{PhysRevB.87.081108} (except possibly in the limit of very strong interactions) that the quasiparticle residue does not effect the leading $L^{d-1} \ln{(L)}$ entanglement term, so we find precise agreement between the dual field theory entropy calculation and the bulk entropy calculation.

There is also the possibility of additional surface terms localized on the bulk minimal surface \cite{2013arXiv1307.2892F}.  In the hard wall geometry such terms will clearly not lead to any logarithmic modification of the area law, so the bulk fermion contribution is indeed the leading correction.

\section{Hyperscaling violation geometries}

We can also consider more general situations.  By including into the above setup an extra scalar field, the dilaton $\phi$, we can construct new types of compressible solutions.  As detailed in Ref. \cite{PhysRevB.85.035121}, the metric can be written in the form
\beq \label{hsvg}
ds^2 = \frac{- f(r) dt^2 + g(r) dr^2 + dx^2 + dy^2}{r^2}
\eeq
for some functions $f$ and $g$.  As above, the asymptotic boundary (field theory UV) is at $r=0$.  First ignoring the bulk fermion $\Psi$ but including the dilaton $\phi$, the solution is roughly divided into two regions, a near boundary AdS-like region near $r=0$ and a deep IR region approaching $r = \infty$.  The deep IR region is called a hyperscaling violation geometry because the metric transforms only up to a conformal factor under rescaling transformations.  The crossover between these two regimes occurs at $r \sim Q^{-1/2}$ where $Q = Q_h$ is the ``hidden" charge density.  When explicit bulk fermions are included, in the fluid approximation they are also found near $r = Q_h^{-1/2}$ where $Q_h = Q - Q_v$, $Q$ is the total charge density, and $Q_v$ is the charge density in the bulk fermions. The precise location of the fluid depends on the details including the dilaton potential, but we argue below that these details have no effect on the entanglement entropy.  Finally, note that in the Thomas-Fermi approximation, the bulk fermions are confined within a finite extent in the $r$ direction.

\begin{figure}
  \centering
  \includegraphics[width=.48\textwidth]{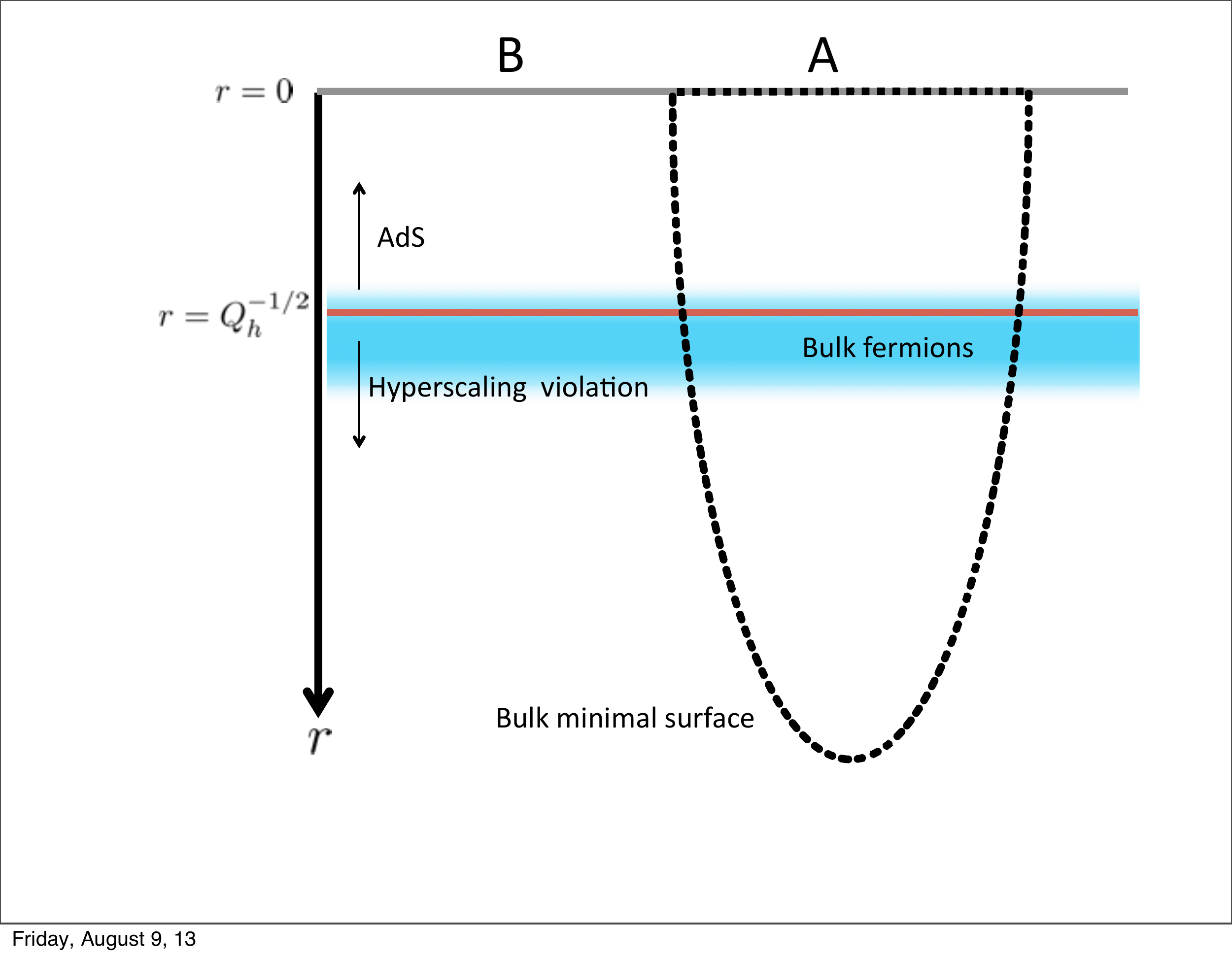}
  \caption{Sketch of a one dimensional slice through the hyperscaling violation geometry with $d=2$.  The near boundary region is AdS-like while the deep IR is of hyperscaling violation form.  The crossover (red line) is roughly at $r\sim Q_h^{-1/2}$ where the bulk fermions (blue region) also reside (for $Q_v \ll Q_h$).  The dashed curve is an exaggerated representation of the bulk minimal surface which hangs down much below all other scales and is approximately a straight cylinder in the region occupied by the bulk fermions.}
  \label{fig:mshv}
\end{figure}

The essential points then mirror the calculations for the hard wall case, see Fig. \ref{fig:mshv}.  In the fluid approximation, there are many bulk fermions and they form a Fermi liquid ground state which may be treated in a hydrodynamic approximation.  At this level of approximation, the quantum state of the fermions still consists of a series of filled bulk Fermi seas.  And just as above, when the linear size $L$ is much greater than $Q_h^{-1/2}$, the bulk minimal surface hangs down straight through the region where the bulk fermions reside.  Thus as far as the bulk fermions are concerned, the entangling surface might as well be a straight cylinder of the form $\partial A \times [0,\infty)$, and the arguments above immediately imply that the bulk entanglement entropy is a sum of many Fermi surface contributions
\beq
\Delta S = \sum_\ell \frac{k_{F,\ell} |\partial A|}{6 \pi} \ln{(k_{F,\ell} L)} + ...
\eeq
where $...$ denotes subleading terms obeying the boundary law.

More generally, there will be some leakage of the bulk fermion wavefunctions outside of the region where the bulk minimal surface is well approximated by a cylinder.  These wavefunction tails are expected to give rise to subleading corrections to the leading $L \ln{(L)}$ result.

We must also consider the possibility of terms localized to the minimal surface.  However, while such terms will effectively modify the value of $G_N$ and hence the coefficient of the classical entropy result, the modification should be essentially the same as that occurring in the zero density state.  This is because the bulk fermions reside only near $r \sim Q_h^{-1/2}$, hence for sufficiently large $A$, the extra effect of the finite density of bulk fermions is negligible relative to the zero density radiative correction.  This result implies that, while the prefactor of the minimal surface contribution is modified, the modification is approximately independent of the state of the bulk fermions.  Hence, just as in Ref. \cite{PhysRevB.85.035121} the coefficient of the ``hidden" $L \ln{(L)}$ contribution is independent of the state of the ``visible" fermions apart from a trivial dependence on the density.  We emphasize that this is a non-trivial feature of the classical geometry which persists in the leading quantum correction.

Below we give some more details of the argument in hyperscaling violation geometries.

\subsection{Minimal surface structure}

In this section we demonstrate our earlier claim that the bulk minimal surface effectively hangs straight down, i.e. may be approximated by the cylinder $\partial A \times [0,\infty)_r$ as far as the finite density of bulk fermions is concerned.  We begin by analyzing the case of a spherical entangling region before giving a more general argument for any shape.

To gain intuition, it is useful to consider first the case when the subregion is a $d$-ball.  For concreteness, we work with the case of $d=2$ spatial dimensions in the field theory.  The disk $A$ is taken to have radius $L$.  We consider the background in Eq. \ref{hsvg} and we parameterize the bulk minimal surface by giving $\rho = \sqrt{x^2+y^2}$ in terms of $r$.  The bulk surface area is then
\beq
|W(A)| = 2 \pi \int_\epsilon^{r^\star} dr \frac{\rho \sqrt{g + (\rho')^2} }{r^2}
\eeq
where $\rho' = \partial_r \rho$ and $r^\star$ is defined by $\rho(r^\star) = 0$.  The equation of motion following from minimizing $|W|$ is
\beq\label{eq:surfeom}
\partial_r \left( \frac{1}{r^2} \frac{\rho \rho'}{\sqrt{g + (\rho')^2}}\right) = \frac{\sqrt{g + (\rho')^2}}{r^2}.
\eeq

We consider in turn the case of pure AdS, $g=1$, and the hyperscaling violation geometry, $g=r^2$.  Our goal is to solve the above differential equation perturbatively near $r=0$ where $\rho(0)=L$.  Set $\rho(r) = c_1 + c_2 r^p + ...$ with $c_1 =L$ and $c_2$ and $p$ to be determined.  For pure AdS ($g=1$), we find to leading order near $r=0$
\beq
c_1 c_2 p (p-3) r^{p-4} = r^{-2},
\eeq
so we have a solution for $p=2$ in which case $c_2 = -1/(2 c_1)$.  Thus the radius $\rho(r)$ of the minimal surface in a fixed $r$ plane only decreases by roughly $r^2/L$ provided $r \ll L$. Thus the bulk minimal surface has almost the same radius as the boundary value provided we are interested in $r \ll L$.  In the hyperscaling case, with $g=r^2$, we look for a solution of the form $\rho(r) = c_1 + c_2 r^p \ln{r}$.  Assuming $p>2$, so that we may neglect $\rho'$ compared to $g$ as $r\rightarrow 0$, we find
\beq
\partial_r \left( c_1 c_2 r^{p-4} (p \ln{r} + 1) \right) = r^{-1}
\eeq
which requires $p=4$ and $c_2 = 1/(4 c_1)$.  Note that even though $c_2$ is positive, the $\ln{r}$ term is negative for small $r$, so $\rho$ still decreases as $r$ increases.  Thus in the hyperscaling violation case, the rate of change of $\rho$ is comparatively even smaller near $r=0$.  Hence the bulk minimal surface will change very little for $r \ll L$.

Now we give a similar calculation for small $r$ and large but generic $\partial A$ that again demonstrates that the shape of $\partial A$ is only distorted by a small amount for any $r \ll L$.  Consider a small section of the bulk minimal surface parameterized by giving $H = x_d$ as a function of $r$ and $x_1, ..., x_{d-1}$.  The infinitesimal area element is then
\bea
&& d |W(A)| \sim  dr dx_1 ... dx_{d-1} \frac{\sqrt{g}}{r^2} \times \cr \nonumber \\
&& \sqrt{1 + \sum_{j=1}^{d-1} (\partial_j H)^2 + g^{-1} (\partial_r H)^2} .
\eea
The equation of motion, specializing to the case of $d=2$, is
\bea
&& \partial_x \left( \frac{\sqrt{g}}{r^2} \frac{\partial_x H}{\sqrt{1 + (\partial_x H)^2 + g^{-1} (\partial_r H)^2}} \right) + \cr \nonumber \\
&& \partial_r \left(\frac{\sqrt{g}}{r^2} \frac{g^{-1} \partial_r H}{\sqrt{1 + (\partial_x H)^2 + g^{-1} (\partial_r H)^2}} \right) = 0.
\eea
Taking $g=r^2$, using the ansatz $H = h(x) + \sigma(x) r^p \ln{r} + ...$ and simplifying we find
\bea
&& \partial_x \left(\frac{1}{r}\frac{\partial_x h}{\sqrt{1+ (\partial_x h)^2}} \right) = \cr \nonumber \\
&& - \partial_r \left(\frac{1}{r^3} \frac{r^{p-1} (p \ln{r} + 1) \sigma}{\sqrt{1+(\partial_x h)^2}} \right).
\eea
So we must have $p=4$ in which case we have a relation for $\sigma$ in terms of $h$ as
\beq
\sigma  = - \frac{ \sqrt{1+ (\partial_x h)^2}}{p} \partial_x \left(\frac{\partial_x h}{\sqrt{1+ (\partial_x h)^2}} \right).
\eeq

Using all these results we see that even for arbitrary $\partial A$, as long as $r \ll L$, the bulk minimal surface in a fixed $r$ plane is close to the same shape as $\partial A$.  It might also be interesting, however, to systematically study the subleading terms arising in this analysis.

\subsection{Including the fluid}

For a large enough entangling region, the minimal surface hangs straight down into the bulk and close to the boundary it has very small dependence on the holographic direction. We verified this both in an AdS geometry as well as in a hyperscaling violation geometry. We now verify numerically that this is also true for the holographic set up we considered in our previous work \cite{PhysRevB.85.035121}. Here the visible fermions are incorporated as a fluid in a geometry that is hyperscaling violating in the IR and AdS in the UV. The fluid description corresponds to the Thomas-Fermi approximation, where the fermion wavefunctions are strongly localized, and has the advantage that it takes into account the backreaction of the visible fermions on the metric \cite{PhysRevD.83.046003}. It follows that in this case there is a crossover region from the hyperscaling violation geometry to the AdS geometry and that there is a backreaction on the metric in the region where the fluid is present. We will see that these two features do not change the expected result, namely that for a large enough entangling region, the minimal surface hangs straight down through the fluid region.

The numerical set up is precisely as in our previous work \cite{PhysRevB.85.035121}  and we refer the reader to section IV.A of that work for details. Here we give a brief summary. The action we consider is the Einstein-Maxwell-dilaton-fluid action,
\bea
\mathcal{L}_{\rm EMDF} &=&  \frac{1}{2 \kappa^2} \Bigl( R  - 2 \left(\nabla \Phi \right)^2   - \frac{V(\Phi)}{L_{\rm AdS}^2} \Bigr)  \nn
& & - \frac{Z(\Phi)}{4 e^2}  F_{\mu\nu}F^{\mu\nu}+p(\mu_{\rm loc}),\nonumber
\eea
where $R$ is the Ricciscalar, $\Phi$ is the neutral scalar dilaton field, with $V(\Phi)$ its potential and $Z(\Phi)$ its coupling to the Maxwell fields. The Maxwell flux $F_{\mu \nu}$ is associated with the vector potential $A_{\mu}$ in the usual way and $p$ is the pressure of the fluid. The pressure is a function of the local chemical potential
\beq
\mu_{\rm loc}=\frac{A_t}{\sqrt{-g_{tt}}}. \nonumber
\eeq
Finally, $\kappa$ is the surface gravity and $L_{\rm AdS}$ the AdS radius. The pressure of the fluid describing fermions with mass, $m$, can be expressed in terms of the energy and charge density
\bea
& &-\hat{p}=\hat{\rho}-\frac{h}{\sqrt{f}} \hat{\sigma}, \nn
& &\hat{\sigma}=\hat{\beta} \int_{\hat{m}}^{\hat{\mu}_{\rm loc}} \epsilon \sqrt{\epsilon^2-\hat{m}^2}d\epsilon, \, \hat{\rho}=\hat{\beta} \int_{\hat{m}}^{\hat{\mu}_{\rm loc}} \epsilon^2 \sqrt{\epsilon^2-\hat{m}^2}d\epsilon, \nonumber
\eea
for $\hat{m}<\hat{\mu}_{\rm loc}$ and zero otherwise. We introduced the dimensionless variables
\bea
p = \frac{1}{L_{\rm AdS}^2 \kappa^2} \hat{p} \,, \qquad \rho = \frac{1}{L_{\rm AdS}^2 \kappa^2} \hat{\rho} \,, \qquad \sigma = \frac{1}{e L_{\rm AdS}^2 \kappa} \hat{\sigma} \,, \nn
\hat{\beta} = \frac{e^4 L_{\rm AdS}^2}{\kappa^2} \frac{1}{\pi^2} \,, \qquad \hat{m^2} = \frac{\kappa^2}{e^2} m^2 \,, \qquad \hat{\mu}_{\rm loc} = \frac{\kappa}{e} \mu_{\rm loc} \,.  \nonumber
\eea
We choose $V(\Phi)$ and $Z(\Phi)$ such that they interpolate between AdS in the UV, where $r\to 0$ and $\Phi \to 0$, and hyperscaling violating in the IR, where $r \to \infty$ and $\Phi \to \infty$. In the UV and IR they thus take the following form respectively,
\bea
\left\{ \begin{array}{ll}
V(\Phi)=-6+2 M_{\Phi}^2 L_{\rm AdS}^2  \Phi^2, \ Z(\Phi)=1 & \textrm{in the UV,}\\
V(\Phi)=-V_0 \exp(\alpha \Phi/3), \ Z(\Phi)=\exp(\alpha \Phi) & \textrm{in the IR.}
\end{array} \right. \nonumber
\eea
For computational convenience we will take the dilaton mass to satisfy $M_{\Phi}^2=-2/L_{\rm AdS}^2$, such that the dual operator $\mathcal{O}$ has scaling dimension $\Delta=2$. Explicitly, we take the following expressions
\bea
V(\Phi) &=& \frac{-V_0}{2 \cosh (\alpha \Phi/3)}+\left(\frac{V_0}{2}-6 \right)(1-\tanh(\alpha \Phi/3)^2) ,\nn
Z(\Phi) &=& \exp(\alpha \Phi), \nonumber
\eea
with $V_0=24 \left(\alpha ^2+6\right)/{\alpha ^2}$ to get the desired dilaton mass in the UV. Finally, without loss of generality we take $\alpha=3$ for computational convenience.

\begin{figure}
  \centering
  \includegraphics[width=.48\textwidth]{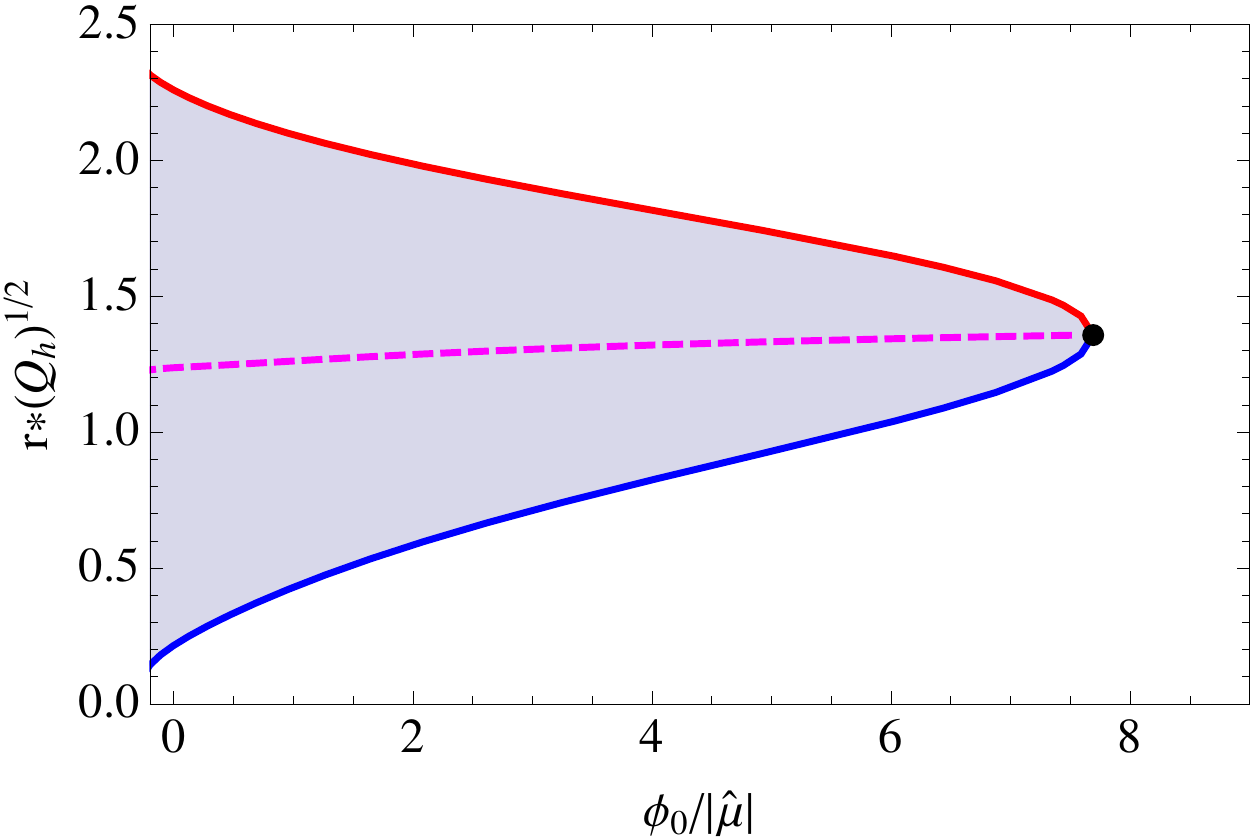}
  \caption{We plot the region, $r_1<r<r_2$, where the fluid resides, with $r_1$ in red and $r_2$ in blue, as a function of the dimensionless parameter, $\phi_0/|\hat{\mu}|$. The dashed magenta line is the center of the region $(r_1+r_2)/2$. All radii are multiplied by a factor $\sqrt{Q_h}$ to obtain dimensionless units. It is clear that the fluid region is centered around $r \sim Q_h^{-1/2}$.}
  \label{fig:fluidradii}
\end{figure}

As we explain in more detail in \cite{PhysRevB.85.035121,0264-9381-29-19-194001} , we can numerically solve the equations of motion for this theory and we find a one parameter family of solutions. The free parameter, $\phi_0/|\hat{\mu}|$, is the dimensionless combination of the boundary chemical potential, $\hat{\mu}$, and the coupling, $\phi_0$, which is the coupling to the relevant operator, $\mathcal{O}$, the operator dual to the dilaton. For large values of $\phi_0/|\hat{\mu}|$ we find that there is no fluid, i.e. $\hat{\mu}_{\rm loc}<\hat{m}$ everywhere in the bulk. It follows that $Q_v=0$ and the total charge, $Q=Q_h+Q_v=Q_h$. This is the fully fractionalized phase. When we dial down $\phi_0/|\hat{\mu}|$, there is a third order transition to the partially fractionalized phase, where the fluid is present in the bulk in a region given by $r_1<r<r_2$. As we argued before, the region is centered around $r \sim Q_h^{-1/2}$, while the width of this region grows with increasing $Q_v$. This can be seen nicely in figure \ref{fig:fluidradii}. We note that it may be possible to shift the location of the fluid region around a bit by choosing a different dilaton potential, however, since there is only one scale in the problem, this cannot lead to any parametric change in the location.

\begin{figure}
  \centering
  \includegraphics[width=0.48\textwidth]{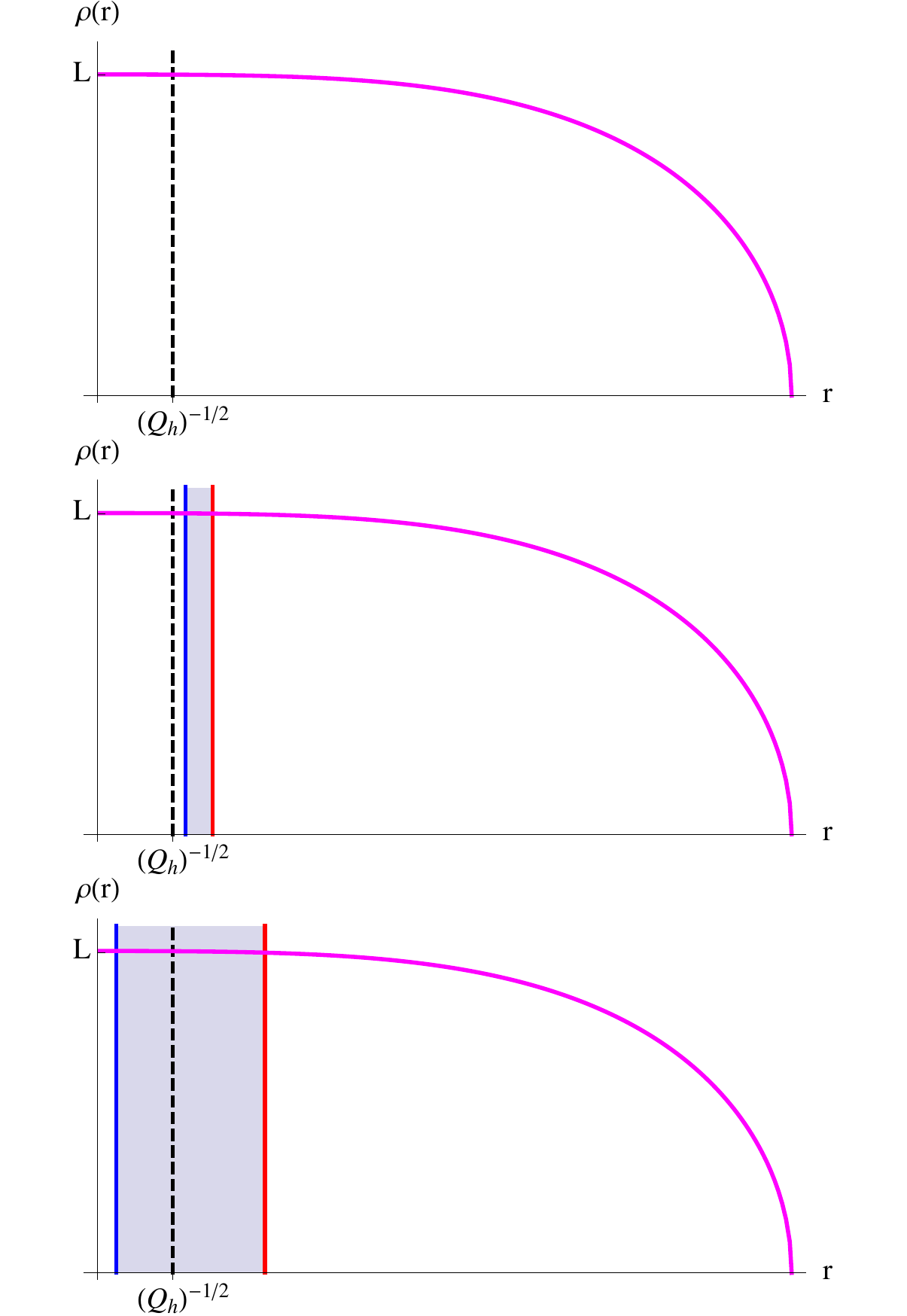}
  \caption{We plot the coordinate $\rho$ of the minimal surface as a function of the holographic direction $r$. The entangling region is a disk of radius $L$ in dimensionless units. From top to bottom, we have $Q_v/Q=0$, $Q_v/Q\sim10^{-5}$ and $Q_v/Q\sim10^{-1}$. The fluid region $r_1<r<r_2$ is indicated by the shaded area, with $r_1$ in red and $r_2$ in blue. We also indicate the location of $r=Q_h^{-1/2}$ by the black dashed line.}
  \label{fig:minsurf}
\end{figure}

To check that the minimal surface hangs straight down through the fluid region, provided the entangling region is large enough, we compute the minimal surface of a disk like entangling region for two values of $\phi_0/|\hat{\mu}|$. We choose $\phi_0/|\hat{\mu}|\approx7.1$, for which $Q_v/Q\sim10^{-5}$, and $\phi_0/|\hat{\mu}|\approx0.12$, for which $Q_v/Q\sim10^{-1}$. Finally, for comparison, we also compute the minimal surface in the fully fractionalized phase, i.e. $Q_v/Q=0$. To compute the minimal surface we solve the equation of motion, (\ref{eq:surfeom}), for the numerically obtain metric. It turns out that stability of the numerics requires solving the equation of motion by shooting out to the boundary from $r^*$, defined by $\rho(r^*)=0$, instead of shooting in from the boundary. The solution of the equation of motion near $r^*$ takes the form
\beq
\rho(r) = g(r^*)r^* \sqrt{r^*-r}+ \dots
\eeq
It is clear from Fig. \ref{fig:minsurf} that neither the crossover between AdS and hyperscaling violation nor the presence of the fluid changes the conclusion that for large enough entangling region the minimal surface hangs straight down in the region where the fluid resides.

\section{Mutual information, Renyi entropy, and subleading oscillatory terms}

In this section we consider more general measures of entanglement and correlation.  Of particular interest is the mutual information $I(A,B)$ between two regions $A$ and $B$ defined by
\beq
I(A,B) = S(A) + S(B) - S(AB).
\eeq
It is also useful to generalize the entanglement entropy to the Renyi entropy $S_n(A)$ defined by
\beq
S_n(A) = \frac{1}{1-n} \ln{\left(\text{tr}(\rho_A^n)\right)}.
\eeq
We now discuss both quantities in the context of hyperscaling violation geometries.

In Ref. \cite{PhysRevB.86.045109} the mutual information of a Fermi gas was computed.  Those results, when applied to the bulk fermion state, imply that in both the hard wall and hyperscaling violation geometries, the bulk fermions give a mutual information which decays as $\ell^{-3}$ in the large $\ell$ limit where $\ell$ is the distance between $A$ and $B$.  The equal time $\Psi$ two point function at separation $\Delta x$ decays as $|\Delta x|^{-3/2}$ and hence the decay of the mutual information is consistent with the bound
\beq
\langle O_A O_B \rangle_c^2 \leq ||O_A ||^2 ||O_B ||^2 I(A,B)
\eeq
where $\langle O_A O_B \rangle_c$ is the connected two point function.

We may also consider the Renyi entropy.  As shown in Ref. \cite{PhysRevB.86.045109}, the bulk Renyi entropy of the bulk Fermi gas also scales like $L \ln{(L)}$ and goes like
\beq
S_{\text{bulk fermion},n} = \sum_{\ell < \ell_0} \frac{1}{2}\left(1 + \frac{1}{n} \right) \frac{k_{F,\ell} |\partial A|}{6 \pi} \ln{(k_{F,\ell} L)} + ...
\eeq
where again $...$ denotes subleading terms.  Assuming the dual fermion two point function, which has a quasiparticle pole, is indicative of a Fermi liquid state in the dual field theory, the Renyi entropy of the dual Fermi liquid also has the same $n$ dependent factor.  Hence again we find agreement between a holographic and a field theory calculation.  However, in the case of the Renyi entropy, the holographic prescription is not really known.  Our calculation here leads us to conjecture that the bulk Renyi entropy across the minimal surface must be involved in the computation of loop corrections to the field theory Renyi entropy.  Nevertheless, it must be emphasized that the minimal surface prescription fails already at the classical level when considering Renyi entropies and it is not known if there is a simple way to compute the leading contribution to the Renyi entropy using properties of an unbranched ($n=1$) geometry (the formal answer is given by the partition function of the branched geometry).

Finally, it is worth noting that among the many subleading terms coming from the bulk entanglement entropy, the presence of bulk Fermi surfaces implies the existence of special oscillating terms tied to the Fermi wavevector \cite{2013PhRvB..87w5112S}.  These oscillating terms are expected on general grounds and using the results of Ref. \cite{2013PhRvB..87w5112S} we see that they have the precise form expected of the dual Fermi liquid state.  Curiously, it seems that here too the quasiparticle residue is irrelevant as confirmed in a solvable model \cite{2012arXiv1209.0769S}.  Friedel oscillations have been obtained holographically in one dimension using monopoles \cite{2012arXiv1207.4208F}, but it remains to be seen if we can access Friedel-like oscillations in correlation functions or entanglement due to hidden Fermi surfaces in higher dimensions.

\section{Discussion}

In this work we have applied to the proposal of Ref. \cite{2013arXiv1307.2892F} to compute quantum corrections to the entanglement entropy of the dual field theory in compressible phases with bulk fermions.  The corrections were shown to violate the boundary law and to conform precisely to the Fermi gas form.  Since the dual fermions are expected to be in a Fermi liquid state, we find agreement between the holographic and dual field theory computations of the entropy.  In terms of the hidden charge density $Q_h$ and the visible charge density $Q_v$ we find an entropy going like
\beq
S \sim s_1 (Q_h^{1/d} L)^{d-1} \ln{(Q_h^{1/d} L)} + s_2 (Q_v^{1/d} L)^{d-1} \ln{(Q_v^{1/d} L)}. \nonumber
\eeq
Crucially, $s_1$ and $s_2$ are independent of any other details of the state, and moreover, $s_2$ may be computed both holographically and in the field theory with perfect agreement.  We also reproduced the detailed shape dependence of the Fermi liquid contribution to the entanglement.  Finally, we studied the mutual information and made some observations about Renyi entropies and oscillating terms.

Based on our analysis, it seems that to obtain non-Fermi liquid behavior in the loop correction, we must have bulk fermions that either exist throughout the whole IR geometry or we must imagine the bulk fermions are themselves in a non-Fermi liquid state, e.g., as may occur in dense neutron stars.  An interesting attempt to study more general back reacted geometries in which the fermions might explore more of the geometry may be found in Ref. \cite{2013arXiv1306.6075A}.

There is also the question of how to interpret the fact that fermions sitting near a definite value of $r$ in the bulk are nevertheless associated with gapless modes.  This observation (and related issues, e.g., bulk Goldstone modes) raises questions about the generality of the radius to RG scale correspondence once loop corrections are considered.  Certainly some modification is expected once the geometry begins to fluctuate. In our case, it seems that the fact that the fermions sit near a definite value of $r$ indicates that they are sharp excitations in the dual field theory.  By comparison, fermions spread throughout the geometry might indicate the lack of a sharp quasiparticle.

\textit{Acknowledgements.} We thank S. Hartnoll and T. Faulkner for feedback on the manuscript.  LH thanks J. de Boer, K. Holsheimer and B. Freivogel for discussions on related topics.
BGS is supported by a Simons Fellowship through Harvard University. This research was supported by the US NSF under Grant DMR-1103860 and
by the John Templeton foundation. 
This research was also supported in part by Perimeter Institute for
Theoretical Physics; research at Perimeter Institute is supported by the
Government of Canada through Industry Canada and by the Province of
Ontario through the Ministry of Research and Innovation.

\bibliography{holo_ee_f_loop}
\end{document}